\documentclass[runningheads]{llncs}

\usepackage[T1]{fontenc}
\def\doi#1{\href{https://doi.org/\detokenize{#1}}{\url{https://doi.org/\detokenize{#1}}}}
\usepackage{graphicx}
\usepackage{listings}
\usepackage{amsmath}
\usepackage{cite}
\usepackage{hyperref}
\begin{document}
\title{Introducing Individuality into Students' High School Timetables\thanks{A. Krystallidis and R. Ruiz-Torrubiano acknowledge financial support from the Research Promotion Agency of Lower Austria (GFF) under project grant FTI21-A-002.}}
%
%
\author{Andreas Krystallidis\inst{1}\orcidID{0009-0007-0183-3038} \and
Rub\'en Ruiz-Torrubiano\inst{1}\orcidID{0000-0001-9314-0739}}
\authorrunning{A. Krystallidis and R. Ruiz-Torrubiano}
%
\institute{IMC Krems University of Applied Sciences, Piaristengasse 1, Krems, 3500, Austria 
\email{andreas.krystallidis@fh-krems.ac.at}\\
\email{ruben.ruiz@fh-krems.ac.at}\\
}
\maketitle              
\begin{abstract}
In a perfect world, each high school student could pursue their interests through a personalized timetable that supports their strengths, weaknesses, and curiosities. While recent research has shown that school systems are evolving to support those developments by strengthening modularity in their curricula, there is often a hurdle that prevents the complete success of such a system: the scheduling process is too complex. While there are many tools that assist with scheduling timetables in an effective way, they usually arrange students into groups and classes with similar interests instead of handling each student individually. In this paper, we propose an extension of the popular XHSTT framework that adds two new constraints to model the individual student choices as well as the requirements for group formation that arise from them. Those two constraints were identified through extensive interviews with school administrators and other school timetabling experts from six European countries. We propose a corresponding ILP formulation and show first optimization results for real-world instances from schools in Germany.

\end{abstract}
\section{Introduction}

As educational systems continuously evolve, crafting close-to-optimal high school timetables continues to pose a major challenge. An important aspect that contributes to this phenomenon is the increasing demand for personalized and flexible learning experiences by all stakeholders of the educational system (policy makers, teachers, students and society in general). This demand results in modular educational systems, where students can choose parts of their own curriculum individually. However, conventional approaches to the creation of timetables can not fulfill the diverse needs of students. This paper addresses the need for innovation in the encoding of constraints for high school timetables, aiming to support the flexibility that modern modular educational systems afford to students.

While traditional formats of constraint encoding do support most of the requirements that schools have on timetables including the possibility to work with individual students instead of classes, they are not able to represent the requirements of each student pursuing diverse individual academic interests in an adaptive manner. In this paper, we present an extension of the XHSTT format~\cite{post_xml_2014} (introduced in the Third International Timetabling competition~\cite{thirdITC}) that incorporates two new constraints, making possible to encode flexible student course choices and class formation requirements while still supporting all previous instances developed for this format. The need for those two new constraint types became apparent through recent research that explored the timetabling demands for schools across central Europe~\cite{ruiztorrubiano2024scheduling} which shows that there is a trend of enabling students to follow their individual interests. Even though this new formulation features students choosing their own respective courses, the problem is still very different from University Course Timetabling Problems~\cite{secondITC, courseTimetabling}, which were already found to be solvable when transformed to the XHSTT format~\cite{fonseca2016modelling}. 

While, our new formulation has some similarities to the student choices featured in the Post Enrolment based Course Timetabling (PE-CTT) Problem, which was first presented in the second track of the ITC 2007~\cite{lewis2007post}, the problem we are modeling here has some key differences: In the PE-CTT Problem students select a set of events that they want to attend without providing alternatives or preferences. It is a hard constraint that they visit all their selected events, while in our newly formulated Constraints it is possible to specify that only a subset of a flexible quantity should be attended. Some additional flexibility is provided in the University Course Timetabling Problem featured at the ITC 2019~\cite{itc2019}. While the students still attend a fixed amount of selected courses, the courses themselves can be split into structured subparts that could be used to model distributions of students to equivalent courses (e.g. Math\_1\_1 and Math\_1\_2). However, to the best of our knowledge, there is no constraint in any Educational Timetabling format that would enable us to model student choices on the individual level in the flexible manner that is desired in modular educational systems. Finally, since we are solving High School Instances and not University Instances, we use almost all the different constraints of XHSTT, so it is much more efficient to extend this format instead of an University Timetabling format that focuses on other qualities that we mostly do not need (e.g. differentiating between different weeks of the semester and complex orderings and structures of events).

We hope that by extending the standard high school timetable format we will be able to spur new original research that adapts and improves methods of automatic timetabling for modular high schools. We support these developments explicitly by providing an ILP formulation that extends one of the state-of-the-art ILP formulations by Kristiansen et al.~\cite{kristiansen2015integer} as well as 18 publicly available instances that include the new constraints. Additionally, we present first results for upper bounds using our ILP.

This paper is structured as follows: In Section 2, we introduce the new constraints as well as how they can be used and provide examples of how to encode various situations that may occur in modular school timetables. In Section 3, we provide a corresponding extension to the well-known ILP formulation by Kristiansen et al.~\cite{kristiansen2015integer}. In Section 4, we describe the new instances in our format and provide some first results. The instances are made publicly available as a benchmark set for modular high school timetabling. Finally, in Section 5, we provide an overview of our findings and describe possible directions of future research.

\section{An Extension to the XHSTT Format}

The XHSTT format~\cite{post_xml_2014} is the most widely used format for encoding the High School Timetabling Problem. It is versatile enough to encode many real-world instances of timetable requirements from various schools around the world accurately. However, in recent years there has been a trend of giving students more choices in what individual courses they wish to attend. While some of the requirements that arise from those choices can be modeled using the existing constraints, others are not supported by the XHSTT format. Since the goal of the XHSTT format is to provide a way to encode timetable requirements in a unified way, we find it important that those recent developments reflect themselves in the form of an extension of the format. This extension should be as small as possible while still being able to accurately encode the new constraints for timetables. Furthermore, we find that everything that can be encoded with the existing set of constraints should still be encoded using them (even when it is in a slightly roundabout way) in order not to put an unnecessary strain upon those who maintain and possibly want to extend existing methods and solutions for solving the High School Timetabling Problem. Finally, it is also important that the format only includes constraints that are actually useful for those in charge of creating the timetables. That is why the constraints proposed in this paper are chosen based on a study~\cite{ruiztorrubiano2024scheduling} where experts across Europe were asked about the challenges they face when creating timetables for high schools. The result of all of those requirements are three new constraint categories that encode student choices, class size requirements and class size balance. Of those three requirements, the class size requirements will be encoded using the existing constraints of the XHSTT format while the other two require one new constraint type each.

\subsection{Student Choices}

From the interviews conducted in the paper by Ruiz-Torrubiano et al.~\cite{ruiztorrubiano2024scheduling} it becomes apparent that many schools offer course choices to students in one form or another, especially in the respective upper cycles. Such choices can range from choosing a general direction (profile) for their studies, which usually results in scheduling all students with a given profile together, to individual course choices subscribing them to specific courses together with other students that made the same election. However, once there is a certain level of modularity it is usually in practice infeasible to schedule the courses in such a way that every student can attend exactly the courses they have selected. To deal with that problem schools have adopted two different methods to manage student course choices. Either they create a timetable first, and students choose lectures that fit into their individual schedules, or students give priorities and/or alternatives for their choices and the timetabler tries to fulfill those preferences to the best of their ability. The first possibility is already supported in XHSTT by simply creating courses with no classroom assignment together with some time preference constraints for said courses. However, the second method requires a possibility to model how many from a pool of courses can be attended on an individual resource level. So for example, a student that wishes to learn another language might choose to attend a Spanish course, if they do not get into that course they would like to learn Italian and if there is also no more room in the Italian course they might want to learn French. It is a hard requirement for that student that they will exactly attend one of those courses. Another student may want to specialize in natural sciences, and they want to attend biology, physics and chemistry eventually while not caring exactly how many of those subjects they will attend in the following year as long as it is at least one. An example for how this constraint may look like can be seen in Listing ~\ref{xml:student:choice}. Additionally, the school might also support that students provide weights (preferences) to their choices. Note that while we designed this constraint to model course choices for students, it can also be used to model teaching preferences, which are also a common theme for many high schools.

All those requirements can be unified into a new constraint type which we call Student Choice Constraint. The constraint has the standard children that all XHSTT constraints share (Id, Name, Required, Weight, CostFunction) the AppliesTo tag consists of Resources and ResourceGroups children. Additionally, the constraint has the child categories EventGroups, Minimum and Maximum.

\lstset{
    language=xml,
    tabsize=3,
    caption=Student Choice Example,
    label=xml:student:choice,
    frame=shadowbox,
    xleftmargin=20pt,
    framexleftmargin=15pt,
    numbers=left,
    numberstyle=\tiny,
    numbersep=5pt,
    breaklines=true,
    showstringspaces=false,
    basicstyle=\footnotesize,
    emph={food,name,price},emphstyle={\color{black}}}
\begin{lstlisting}
<StudentChoiceConstraint Id="StudentChoice_ST_Bob">
    <Name>StudentChoice_ST_Bob</Name>
    <Required>true</Required>
    <Weight>50</Weight>
    <CostFunction>Linear</CostFunction>
    <AppliesTo>
        <ResourceGroups/>
        <Resources>
            <Resource Reference="ST_Bob"/>
        </Resources>
    </AppliesTo>
    <EventGroups>
        <EventGroup Reference="Biology_10"/>
        <EventGroup Reference="Physics_10"/>
        <EventGroup Reference="Chemistry_10"/>
    </EventGroups>
    <Minimum>1</Minimum>
    <Maximum>3</Maximum>
</StudentChoiceConstraint>
\end{lstlisting}

\begin{itemize}
    \item \textbf{AppliesTo:} Each resource that is either part of the Resources child or is part of a resource group, which is mentioned in the ResourceGroups child, is a point of application for this constraint. 
    \item \textbf{EventGroups:} All Event Groups mentioned in this child are relevant to the constraint.
    \item \textbf{Minimum:} Each resource this constraint applies to has to attend at least Minimum Event Groups from the relevant Event Groups for this constraint.
    \item \textbf{Maximum:} Each resource this constraint applies to can, at most, attend Maximum Event Groups from the relevant Event Groups for this constraint.
\end{itemize}

The deviation of the constraint is described as follows: For each resource part of the Resources and ResourceGroups the deviation is equal to the number of Event Groups that the resource attends, which exceed the Maximum or fall short of the Minimum. Note that we define attendance as visiting any subevent of a given Event Group.

This constraint can also be used to encode weighted preferences. Imagine a student who wants to attend one out of 3 courses $A$, $B$, and $C$ but has a preference order of $A \succ B \succ C$. First, a hard constraint can be added where the EventGroups child contains all 3 courses, and the Minimum and Maximum are set to 1. We then add a soft constraint containing only Events $A$ and $B$ and another constraint containing only event $A$. Again we set the Minimum and Maximum to 1 for both soft constraints. Depending on the weights of the Soft constraints we can now adjust the importance of the student getting his first or second choice.

\subsection{Class Sizes}

Whether it is due to room limitations, pedagogic restrictions, or legal reasons (supervision duties), schools usually have limits on how large the classes for each course can be at most. In a system without student choices, this is usually enforced when the classes are put together before scheduling the individual lessons. Events where the classes are split and mixed are often modeled using one main Event (to which the whole class is assigned) and multiple subevents that are all linked to the main event under the assumption that none of the students attend multiple of those subevents. This works, for example, if one wants to split all students from one class level into two math groups, and every student has to choose a second foreign language. However, this method quickly becomes more complex the more individual the student choices become since it most likely won't be possible to build sets of subjects that have no student overlaps while also not creating many idle periods in student timetables and giving all of the students their preferred subjects, which means that there is a need to find some optimal balance between those constraints. Usually there are restrictions that don't allow for any idle time in student timetables during certain periods, while other periods are more flexible (e.g., in the afternoon). Building the classes as part of the optimization problem allows the solver to find which classes to group dynamically based on when it schedules them, also taking into account how important it is to fulfill each individual student's choice. We can model those restrictions using the existing XHSTT constraints, which makes it easier for existing approaches to adapt to those changes. In the following, we describe exactly how to model this class size problem because the translation into XHSTT constraints is not completely trivial. However, first, we want to recap how the XHSTT format works on a high level.

An instance in the XHSTT format consists of Times, Resources, Events, and Constraints. The Events have a duration, which specifies how many time slots must be assigned to them. How exactly those times are distributed over the week is specified as part of the Constraints. We say that each block of consecutive time slots that is part of the final schedule builds a SolutionEvent (or subevent). Each of those SolutionEvents specifies some resource requirements that can either come in the form of a fixed AssignedResource or a flexible UnassignedResource. In the case of an UnassignedResource, there are often some constraints that restrict the pool of possible concrete resource assignments. Finally, the set of Constraints also consists of other types of Constraints that further impose limits on when, how often, and in what constellations Events, Resources, and groups of Events and Resources are scheduled. 

With that basic understanding of the XHSTT format, we can now get into how we modeled the class sizes. First of all, we assume that three resource types exist (but it is possible to arbitrarily add more): Teachers, Rooms, and Individual Students. If fixed classes still exist for certain courses they can easily be modeled by assigning all students of that class directly to that course. For those courses that should be built by the solver, we create three event types: 
\begin{enumerate}
    \item  One main event that will be used in all constraints that handle the time assignment of the lessons.  This main event will also be used to either directly assign a room and teacher or model the resource preferences for those two resource types. Any restrictions on how the event should be split and distributed over the week will also be applied here

\item As a next step, we create one event for each student resource that is required to fulfill the minimum student number of the course $s_{min}$ (e.g., if it needs at least 10 students to build a language class we create 10 events). They only have one student event resource which is usually not preassigned (can optionally be preassigned to a specific student if attendance is mandatory). We also need a hard Assign and Prefer Resource Constraint so that only those students who choose the course can be assigned and all events must have a student assigned. 

\item Next, we need to add a hard Avoid Split Assignments constraint for each Event (modeled with an Event Group that only contains one student Event) which ensures that two subevents of the same Event can't have different student assignments. 

\item Afterwards, we add a hard Link Events Constraint to the main Events so that all subevents must have the exact same time assignments as the main Event. Through hard Avoid Clashes Constraints on the individual students this also guarantees that each subevent must have a different student assigned. We will henceforth call Events of this type "minimum requirement events".

\item As a final step, we create Events that are very similar to the previous ones but contain those students that are optional from the Event perspective. We create a total of $s_{max} - s_{min}$ events of this type. All constraints are the same except that we do not use any Assign Resource Constraints since it is fine if no students are assigned to the Event (The Prefer Resource constraints have to stay so that if a student is assigned it must be one that chose the class). We will henceforth call events of this type "maximum requirement events"

\end{enumerate}

Note that it would be possible to combine the student requirement events into a single event from a modeling perspective (using a separate Role for each student). We decided against this approach in case that some existing solvers might enumerate all combinations of resource assignments for each Event, which would lead to an exponential amount of such combinations. In all other aspects the approaches are to the best of our knowledge equivalent, except that it would be possible to directly quantify over $b$ variables in the two new proposed constraint types. However, quantifying over EventGroups instead of individual Events gives the advantage that we can use more "high level" constraints, for example if we go back to the example provided in Listing \ref{xml:student:choice} Biology\_10, Physics\_10 and Chemistry\_10 could each be an EventGroup that represents multiple courses (e.g. Biology\_10\_1, Biology\_10\_2 and Biology\_10\_3). We would then add three more Student Choice Constraints (one for each subject) for ST\_Bob each with a Minimum set to 0 and Maximum set to 1. The result is that on the high level Bob will visit between 1 and 3 of his selected choices and on the lower level he will be assigned to exactly one course corresponding to the assigned subjects.

\subsection{Class Size Balance}

Sometimes one subject is taught in multiple courses handling exactly the same school material because the number of students is too big for one single classroom and teacher. One way to handle this using existing constraints is to simply use two teachers and rooms for the class while also scaling the student requirements. However, this has the restriction that both of those courses would need to happen in parallel, which takes away some flexibility, especially when the student schedules are very individual. 

A better alternative would be to have two completely separate courses. With the help of Student Choice constraints, we can then model that a student can or must attend one of them. However, this could result in the undesirable property of possibly very unbalanced class sizes (e.g., two math classes, one with the bare minimum assignment of 10 students while the other is fully booked with 30 students). To prevent this, we introduce Balance Class Size constraints that, aside from the standard children that all XHSTT constraints share (Id, Name, Required, Weight, CostFunction), have the tags AppliesTo with the child EventGroups, Role and MaximumDifference. An example for how this constraint would be modeled in the case of two equivalent math classes can be found in Listing \ref{xml:balance:class}.

\lstset{
    language=xml,
    tabsize=3,
    caption=Balance Class Size Example,
    label=xml:balance:class,
    frame=shadowbox,
    xleftmargin=20pt,
    framexleftmargin=15pt,
    numbers=left,
    numberstyle=\tiny,
    numbersep=5pt,
    breaklines=true,
    showstringspaces=false,
    basicstyle=\footnotesize,
    emph={food,name,price},emphstyle={\color{black}}}
\begin{lstlisting}
<BalanceClassSizeConstraint Id="BalanceClassSize_Math_5a">
    <Name>BalanceClassSize_Math_5a</Name>
    <Required>false</Required>
    <Weight>1</Weight>
    <CostFunction>Linear</CostFunction>
    <AppliesTo>
        <EventGroups>
            <EventGroup Reference="Math_1_5A"/>
            <EventGroup Reference="Math_2_5A"/>
        </EventGroups>
    </AppliesTo>
    <MaximumDifference>2</MaximumDifference>
    <Type>Student</Type>
</BalanceClassSizeConstraint>
\end{lstlisting}

\begin{itemize}
    \item \textbf{AppliesTo:} Each Event Group that is mentioned in the EventGroups child is relevant for this constraint. 
    \item \textbf{MaximumDifference:} An integer that sets a limit of how much difference between the number of assigned resources can be between the Event Groups without causing a deviation.
    \item \textbf{Type:} Optional child. If a Type is given only resources with this type are counted towards the assigned resources of each event.
\end{itemize}

The deviation of this constraint is described as follows: For each Event Group that is part of the EventGroups child, the deviation is calculated as the Maximum Difference to the Event Group (part of the Event Groups Child) with either the most or the least assigned resources (depending on which difference is higher) minus the allowed MaximumDifference.

\section{ILP Formulation}

\label{formulation}
In this section we introduce an ILP model that can be used to solve an instance of our extension to the High School Timetabling Problem. For this purpose we will extend the formulation from Kristiansen et al.~\cite{kristiansen2015integer} by our two new constraints as well as the relevant variables and linkings. In this section, we will only describe the new constraints. The full ILP formulation can be found in the Appendix. Note that for the moment our new instances and format only support linear and quadratic cost functions.

\subsection{Sets}

First, we introduce some sets of entities relevant to the extended modular XHSTT problem which are the same as used by the model for the original problem~\cite{kristiansen2015integer}.

\begin{flalign}
    & t \in T & \text{ordered set of times} \\
    & tg \in TG & \text{set of time groups} \\
    & r \in R & \text{set of resources} \\
    & e \in E & \text{set of events} \\
    & eg \in EG & \text{set of event groups} \\
    & er \in e & \text{set of event resources of an event } e \\
    & se \in e & \text{set of subevents of an event } e \\
    & c \in C & \text{set of constraints} \\
    & p \in c & \text{points of application of constraint } c \\
    & d \in p & \text{deviations of a point of application } p \\
    & i \in I & \text{possible deviation values of deviations } d \\
    & j \in J & \text{possible deviation sum values at points } p \\
    & T_{se, t}^{\text{start}} & \text{possible start times for se that occupy } t
\end{flalign}

\subsection{Further Notation}

Next, we need some further notation to express some of the constraints.

\begin{flalign}
    & r_D & \text{dummy-resource with meaning no resource assigned} \\
    & t_D & \text{dummy-time with meaning no time assigned} \\
    & D_e & \text{duration of event } e \\
    & D_{se} & \text{duration of subevent } se \\
    & e \in c & \text{constraint } c \text{ applies to event } e \\
    & r \in c & \text{constraint } c \text{ applies to resource } r \\
    & eg \in c & \text{constraint } c \text{ applies to event group } eg \\
    & w_c & \text{weight of constraint } c \\
    & \rho(t) & \text{index of time } t \text{ in ordered set } T \\
    & PA_{er} & \text{is 1 if event resource } er \text{ has a preassigned resource otherwise 0} \\
    & \overline{B}_c & \text{upper limit of constraint } c \\
    & \underline{B}_c & \text{lower limit of constraint } c
\end{flalign}

\subsection{Variables}

Compared to the formulation by Kristiansen et al.~\cite{kristiansen2015integer}, we added the variables $b$ and $c$ which are required to model if a resource is participating in Events or Event Groups.

\begin{flalign}
    & x_{se, t, er, r} & \text{binary, indicates if } se \text{ starts at } t \text{ and } r \text{ is assigned to } er \\
    & y_{se, t} & \text{binary, indicates that } se \text{ starts at } t \\
    & v_{t, r} & \text{integer, indicates how often } r \text{ is used at } t \text{ by any } se \\
    & w_{se, er, r} & \text{binary, indicates if } se \text{ is assigned } r \text{ for } er \\
    & b_{e, r} & \text{binary, indicates if } r \text{ is assigned to any } se \text{ of } e \\
    & c_{eg, r} & \text{binary, indicates if } r \text{ is assigned to any } e \text{ of } eg \\
    & s_{c, p, d} & \text{integer, deviation } d \text{ at point of application } p \text{ of } c \\
    & s_{c, p, d, i} & \text{binary, indicates that } d \text{ has value } i \text{ at } p \text{ of } c \\
    & u_{c, p, j}^{\text{SquareSum}} & \text{binary, indicates that the sum of deviations at } p \text{ is } j \\
    & u_{se} & \text{binary, indicates whether } se \text{ is active or not} \\
    & q_{r, t} & \text{binary, indicates if } r \text{ is busy at } t \\
    & p_{r, tg} & \text{binary, indicates if } r \text{ is busy at some } t \text{ in } tg 
\end{flalign}

\subsection{Functions}

We also need to introduce some functions that will mainly be used to express the objective value of the problem.

\begin{flalign}
    & f(s_{c, p, d}) = w_c \cdot \text{CostFunction}(s_{c, p, d}) & \text{cost of constraint } c \\
    & CF^{\text{Sum}} = \sum_{p \in c, d \in p} s_{c, p, d} & \text{Sum cost function} \\
    & CF^{\text{SumSquare}} = \sum_{p \in c, d \in p, i \in I} i^2 \cdot s_{c, p, d, i} & \text{SumSquare cost function} \\ 
    & CF^{\text{SquareSum}} = \sum_{p \in c, j \in J} j^2 \cdot u_{c, p, j}^{\text{SquareSum}} & \text{SquareSum cost function}
\end{flalign}

Some constraints have an upper limit and a lower limit. In this case we define the value of deviation $V$ using the function $U_{\underline{B}_c, \overline{B}_c}V$ as follows:

\begin{flalign}
    & s \geq U_{\underline{B}_c, \overline{B}_c}V \rightarrow 
    \begin{cases}
        s \geq V - \overline{B}_c\\
        s \geq \underline{B}_c - V
    \end{cases} 
    & \text{deviation with upper and lower limit}
\end{flalign}

\subsection{Updated Objective Function}

The objective function consists of the sum of all cost functions of individual constraints. We can also split this objective function into separate values $z_{hard}$ and $z_{soft}$ to denote the costs of hard and soft constraints respectively. Compared to the objective function by Kristiansen et al.~\cite{kristiansen2015integer}, we simply extended the function by adding the terms describing the deviation of our new constraint types.

\begin{equation}
\begin{split}
    \min z = & f(s_{c, er}^{\text{assignres}}) + f(s_{c, er}^{\text{assigntime}}) + f(s_{c, e}^{\text{spliteventamount}} + s_{c, e}^{\text{spliteventdur}}) \\
    & + f(s_{c, e, er}^{\text{distsplitevent}}) + f(s_{c, er}^{\text{preferres}}) + f(s_{c, e}^{\text{prefertime}}) + f(s_{c, eg}^{\text{avoidsplit}}) \\
    & + f(s_{c, eg, tg}^{\text{spreadevent}}) + f(s_{c, eg, t}^{\text{linkevent}}) + f(s_{c, r, t}^{\text{avoidclashes}}) + f(s_{c, r}^{\text{unavailabletimes}}) \\
    & + f(s_{c,r}^{\text{idletimes}}) + f(s_{c, r}^{\text{clusterbusy}}) + f(s_{c, r, tg}^{\text{limitbusy}}) + f(s_{c, r}^{\text{limitworkload}}) \\
    & + f(s_{c, eg}^{\text{balancesize}}) + f(s_{c, r}^{\text{studentchoice}})
\end{split}
\end{equation}

\subsection{Constraints}

\subsubsection{Added General Constraints}
\hfill \break
\hfill \break
The model needs several linking constraints and other general constraints to make the variables express the above-described properties. We will only describe those general constraints that were added to the model of Kristiansen et al.~\cite{kristiansen2015integer} the remaining constraints can be found in the Appendix.

The following new constraints link variables $w_{se, er, r}$ and our new variables $b_{e, r}$:
\begin{flalign}
    & w_{se, er, r} \leq b_{e, r} & \forall e \in E, se \in e, er \in e, r \in R \\
    & \sum_{se \in e} w_{se, er, r} \geq b_{e, r} & \forall e \in E, er \in e, r \in R
\end{flalign}

To link our new variables $c_{eg, r}$ to $b_{e, r}$ we need two more new constraints:
\begin{flalign}
    & b_{e, r} \leq c_{eg, r} & \forall eg \in EG, e \in eg \\
    & \sum_{e \in eg} b_{e, r} \geq c_{eg, r} & \forall eg \in EG
\end{flalign}

\subsubsection{Balance Class Size Constraint}
\hfill \break
\hfill \break
\textbf{Applies to:} Event Groups \\
\textbf{Point-of-application:} Event Group \\
We use the parameter $\overline{B}_{c}$ to denote the maximum class size difference specified in constraint $c \in \overline{C}$.
The role parameter is optional and denotes that only resources of a specific role in the event should be considered
We use the variable $mr_{c, eg}$ to denote the number of resources allocated to the specified Event Group
\begin{flalign}
    & \sum_{\substack{e \in eg, er \in e, \\ r \in er, type_{r} = type_{c} \backslash \{r_D\}}} c_{eg, r} = mr_{c, eg} & \forall c \in \overline{C}, eg \in \overline{C} \\
    & mr_{c, eg} - mr_{c, eg2} - \overline{B}_{c} \leq s_{c, eg}^{\text{balancesize}} & \forall c \in \overline{C}, eg \in c, eg2 \in c, eg \neq eg2 \\
    & mr_{c, eg2} - mr_{c, eg} - \overline{B}_{c} \leq s_{c, eg}^{\text{balancesize}} & \forall c \in \overline{C}, eg \in c, eg2 \in c, eg \neq eg2
\end{flalign}

\subsubsection{Student Choice Constraint} 
\hfill \break
\hfill \break
\textbf{Applies to:} Resources \\
\textbf{Point-of-application:} Resource\\
We use parameters $\underline{B}_{c}$ and $\overline{B}_{c}$ to denote the minimum and maximum values specified in constraint $c \in \overline{C}$.
\begin{flalign}
    & U_{\underline{B}_c, \overline{B}_c} \sum_{eg \in c, e \in eg, er \in e, r \in er, r = r_c} c_{eg, r} \leq s_{c, r}^{\text{studentchoice}} & \forall c \in \overline{C}, r_c \in c
\end{flalign}

\subsection{Model Size Reductions}

Using the model described above without any additions results in models that are too big to handle on our computing cluster with 64 GB of RAM for most of the instances. For that reason, we eliminated some variables that would never be used for an acceptable solution. Our variable eliminations consist of the following list:

\begin{itemize}
    \item We eliminated all variables $x_{se, t, er, r}$ and $w_{se, er, r}$ for students $r$ that did not select an Event $e$, $se \in e$. This means that we do not permit solutions where students who did not select an event are assigned to one of its subevents (which is supported by practice).
    \item We only generate subevents with a feasible duration if there is a hard split events constraint restricting the duration and/or amount of generated subevents.~\cite{GOAL} This is the same technique that was used by Fonseca et al.~\cite{ILPTechnique2} to reduce model sizes.
\end{itemize}

\section{Evaluation}

\subsection{Instances}

As a first benchmarking set\footnote{\url{https://github.com/IMC-UAS-Krems/modularXHSTT}}, we chose 18 high schools with modular school systems from 6 different federal states in Germany. Note that the secondary educational systems in Germany can vary greatly depending on the particular federal state~\cite{ruiztorrubiano2024scheduling}, which makes this set more diverse than instance groups from most other countries. The original anonymized instances were provided by Untis GmbH\footnote{\url{https://www.untis.at}}, an Austrian company that specializes in software that assists schools with their various scheduling problems. We implemented methods to automatically translate their format for encoding constraints to the new extended XHSTT format, which enables us to provide many more instances in the future (Untis collaborates with over 26,000 schools worldwide). However, it is important to note that the XHSTT instances are not a one to one match semantically with the original instances provided by Untis. This is due to the complex nature of the Untis specification that uses a lot of empirical experience to evaluate timetables on factors that can't be represented in a standardized format. We still managed to achieve an extended XHSTT formulation that matches the Untis formulation closely. Some statistics of the instances can be found in Tables~\ref{basicInfo} and~\ref{modularInfo}. Table~\ref{basicInfo} describes how many resources of each resource type are used in each instance. Note that we do not include the minimum/maximum requirement events in the event count since we categorize them as part of the original event and they will always be scheduled together. However, we list the number of requirement events as well as other quantifiable properties that describe the modularity of each instance in Table~\ref{modularInfo}. Note that the amount of requirement events is equal to the number of student assignments that can (but don't necessarily have to) happen to modular events. Table \ref{modularInfo} lists how many Student Choice and Balance Class Size constraints each instance uses. The column Modular Events describes the cardinality of the subset of events that have minimum and/or maximum requirement events associated with them. However, there are many more factors that can have an impact on how complex the resulting instance will be. One factor of complexity is the number and type of constraints from the original XHSTT problem definition. Another factor that has a significant impact is the size of the student pool that is feasible for each requirement event. An instance will be much harder if it features some events where a high percentage of the total amount of students wants to participate in certain events. It is also noteworthy that compared to most other benchmark instances of the original XHSTT problem, this instance set is more restrictive on the possible times for events and the available times for students and teachers. Specifically, there are constraints restricting how many primary subjects a student can attend per day/in a row, how many lessons a teacher may teach in a row without a break, minimum and maximum amounts of idle times per week for teachers, global constraints that define a time-window when a lunch break can/must happen for both students and teachers and hard restrictions that allow no student idle times in the hours before lunch. We made sure to choose schools of varying sizes and proportions of modular events so that it will be possible in the future to find out where the complexities of the problem lie. We also plan to extend this set of instances in the future with other modular schools from across Europe to cover more of the possible instance space.

\begin{table}
\caption{Amount of resources present in each of the new instances (by resource type).}\label{basicInfo}
\begin{center}
\begin{tabular}{|l|c|c|c|c|c|}
\hline
Instance & Events & Students & Classes & Teachers & Rooms \\
\hline
GermanyRHPF1 & 1430 & 143 & 141 & 142 & 208 \\
GermanyHAMB1 & 554 & 847 & 10 & 104 & 80 \\
GermanyNRWE1 & 627 & 118 & 52 & 185 & 186 \\
GermanyHAMB2 & 636 & 215 & 43 & 106 & 73 \\
GermanyNRWE2 & 329 & 252 & 16 & 40 & 69 \\
GermanyRHPF2 & 414 & 134 & 25 & 88 & 74 \\
GermanyRHPF3 & 454 & 167 & 25 & 92 & 84 \\
GermanyNRWE3 & 226 & 252 & 16 & 40 & 69 \\
GermanyNRWE4 & 1045 & 331 & 112 & 185 & 310 \\
GermanyRHPF4 & 545 & 894 & 0 & 83 & 106 \\
GermanySAAR1 & 428 & 169 & 25 & 75 & 65 \\
GermanyRHPF5 & 375 & 182 & 21 & 75 & 61 \\
GermanyRHPF6 & 526 & 360 & 30 & 107 & 67 \\
GermanyBAWU1 & 832 & 157 & 76 & 201 & 176 \\
GermanyBAWU2 & 976 & 272 & 59 & 122 & 173 \\
GermanyBAWU3 & 241 & 228 & 22 & 92 & 71 \\
GermanyBAWU4 & 762 & 205 & 24 & 106 & 70 \\
GermanyHESS1 & 705 & 181 & 49 & 177 & 225 \\
\hline
\end{tabular}
\end{center}
\end{table}

\begin{table}
\caption{Quantifyable properties describing the modularity of the new instances.}\label{modularInfo}
\begin{center}
\begin{tabular}{|l|c|c|c|c|}
\hline
Instance & Choice Constraints & Balance Constraints & Modular Events & Requ. Events\\
\hline
GermanyRHPF1 & 2378 & 34 & 142 & 2440 \\
GermanyHAMB1 & 911 & 42 & 46 & 970 \\
GermanyNRWE1 & 1254 & 60 & 73 & 1265 \\
GermanyHAMB2 & 2053 & 62 & 112 & 2206 \\
GermanyNRWE2 & 2408 & 56 & 106 & 2893 \\
GermanyRHPF2 & 1476 & 40 & 93 & 1385 \\
GermanyRHPF3 & 1718 & 50 & 109 & 1892 \\
GermanyNRWE3 & 1677 & 35 & 73 & 1942 \\
GermanyNRWE4 & 3433 & 79 & 179 & 3828 \\
GermanyRHPF4 & 3984 & 100 & 241 & 4063 \\
GermanySAAR1 & 1421 & 57 & 81 & 1596 \\
GermanyRHPF5 & 1904 & 41 & 124 & 2622 \\
GermanyRHPF6 & 2567 & 21 & 131 & 3255 \\
GermanyBAWU1 & 1452 & 45 & 76 & 1764 \\
GermanyBAWU2 & 3743 & 84 & 176 & 4112 \\
GermanyBAWU3 & 2473 & 75 & 162 & 2496 \\
GermanyBAWU4 & 2309 & 61 & 160 & 2347 \\
GermanyHESS1 & 2947 & 24 & 47 & 1189 \\
\hline
\end{tabular}
\end{center}
\end{table}

\subsection{ILP Evaluation}

Based on the size of the new instances and the added complexity from the new constraints we expect the new instances to be more difficult to solve than previous benchmark sets for the XHSTT. Previous experiments~\cite{kristiansen2015integer} using ILP as an exact method have shown that the problem is still too challenging for modern ILP solvers when using bigger instances. Therefore, we expect to only produce weak upper bounds using the above-described ILP model. Nevertheless, we find it important to provide first results for the newly introduced problem, which should show whether the problem is trivial to solve or not. While employing ILP as an exact method might not yield practical solutions directly, past research suggests its potential when integrated into metaheuristic or matheuristic approaches for tackling the High School Timetabling Problem~\cite{ILPTechnique1, ILPTechnique2}.

Table~\ref{ILPruns} displays the outcomes of executing the 18 instances, with a Memory Limit set at 64 GB, on an AMD EPYC 7252 processor utilizing the commercial ILP solver Gurobi version 10.0.1. Each instance was allotted a time limit of 6 hours for computation. This time frame notably exceeds the 1000-second constraint imposed during the ITC2011 competition. The objective value is split into two components of the form (hard, soft) constraint deviations.

It's worth noting that in practical scenarios, schools typically face fewer time constraints when devising their timetables. They are often willing to invest several hours, or even days, in computational time without significant pressure. However, it's crucial to strike a balance between computational resources and research accessibility. Excessive resource consumption could potentially limit accessibility to the problem, which runs counter to the goal of fostering open research.

\begin{table}
\caption{ILP results and statistics on 6-hour run.}\label{ILPruns}
\begin{center}
\begin{tabular}{|l|c|c|c|}
\hline
Instance & Variables & Constraints & Objective Value \\
\hline
GermanyRHPF1 & 9007909 & 67145239 & --- \\
GermanyHAMB1 & 22276651 & 61355937 & --- \\
GermanyNRWE1 & 5282717 & 22736030 & (152422, 2330) \\
GermanyHAMB2 & 6268589 & 19983356 & ---  \\
GermanyNRWE2 & 12183297 & 41285751 & ---  \\
GermanyRHPF2 & 3013943 & 9066705 & (125430, 4663)  \\
GermanyRHPF3 & 6422201 & 21112371 & (157904, 86000)  \\
GermanyNRWE3 & 8581837 & 30750272 & (75295, 206299)  \\
GermanyNRWE4 & 23744735 & 88931169 & --- \\
GermanyRHPF4 & 42657262 & 87422983 & ---  \\
GermanySAAR1 & 4695284 & 14566661 & (142059, 23465)  \\
GermanyRHPF5 & 5870336 & 14030401 & (117175, 379454)  \\
GermanyRHPF6 & 20876398 & 53277518 & ---  \\
GermanyBAWU1 & 7567498 & 27892830 & (231653, 70076)  \\
GermanyBAWU2 & 11729612 & 31872156 & (257726, 348709)  \\
GermanyBAWU3 & 8464935 & 18952959 & (81008, 3416)  \\
GermanyBAWU4 & 7740808 & 19141384 & (193313, 8665)  \\
GermanyHESS1 & 5737775 & 29194865 & ---  \\
\hline
\end{tabular}
\end{center}
\end{table}

The experiment reveals the complexity inherent in the problem, suggesting that exact methods may struggle to provide satisfactory solutions. Out of the 18 instances studied, solutions were only found for 10.

Furthermore, the integral solutions obtained from the experiment did not meet the criteria necessary for a viable school timetable. Despite leveraging Gurobi's optimization capabilities, the solutions fell short, underscoring the intricacies of the problem and the limitations of current methodologies.

Additionally, Gurobi's inability to provide lower bounds within the designated timeframe prevents us from assessing the optimality gap. However, there is potential for progress as the schools that provided the instances have successfully created their own timetables. In general it is not always possible to find feasible solutions for the requirements encoded by schools. Untis deals with this problem by either leaving some hours unscheduled or reporting the problems with the final timetable that will then be manually resolved by the administrator, which then has to decide which hard constraints can be softened. Nevertheless those solutions may enable us to establish tighter upper bounds in the future, enhancing our understanding of the problem and potentially guiding optimization strategies.

\section{Conclusion and Future Work}

In this paper, we proposed an extension to the XHSTT format to address the constraints that are present in the ever-growing number of modular high schools. We explained the reasons that make those changes meaningful and how exactly they can be incorporated into the XHSTT format. Furthermore, we provided 18 new real-world instances from modular high schools of different regions in Germany together with an ILP model that can be used to find feasible schedules. Our experiments showed that using our ILP model as an exact method for finding solutions is not very effective, and even after 6 hours of runtime, it could only find solutions that are nowhere near satisfactory. However, based on the data from previous benchmarks those results were expected and should not discourage us from finding more efficient methods for creating timetables for modular high schools. We believe that the real strength of the presented ILP will be revealed once it is used as part of a heuristic approach like Large Neighborhood Search (LNS).

In future work, we want to use different exact methods, like a SAT solver to inspect if the problem is really as hard as it seems to be or if ILP is simply not the right approach for this problem. There has already been previous work into possible cuts for the original ILP model and we plan to look into possible new cuts for our extension as well. It will also be interesting to see how the various heuristics developed for the original High School Timetabling Problem perform on this extension and if new heuristics can be found that might work even better for this extension. Specifically, we want to look into a more adaptive LNS-based approach to see if methods from Reinforcement Learning can be used to improve the process of finding good schedules. We will support those developments by providing more benchmark instances from all across Europe together with a publicly available tool for validation. Finally, we will also compare future findings with the timetables produced for actual schools to see if there is any gap between theory and practice.

\bibliographystyle{splncs04}
\bibliography{literature}
%

\section*{Appendix}

\subsection*{Complete ILP Formulation}

The variables, sets, functions and general further notation of the model can be found in Section \ref{formulation}. Here we provide the full set of constraints used in the model (with some repetition from Section \ref{formulation} to avoid confusion), as well as some further additions to guarantee exact objective values.

\subsubsection{Constraints}

\paragraph{General Constraints}
Link variables $s_{c, p, d}$ and $s_{c, p, d, i}$:
\begin{flalign}
    & \sum_{i \in I} i * s_{c, p, d, i} = s_{c, p, d} & \forall c \in C, p \in c, d \in c
\end{flalign}
Only one deviation indicator can be set per deviation:
\begin{flalign}
    & \sum_{i \in I} s_{c, p, d, i} = 1 & \forall c \in C, p \in c, d \in c
\end{flalign}
Link variables $s_{c, p, d}$ and $u_{c, p, j}^{\text{SquareSum}}$:
\begin{flalign}
    & \sum_{j \in J} j * u_{c, p, j}^{\text{SquareSum}} = \sum_{d \in p} s_{c, p, d} & \forall c \in C, p \in c
\end{flalign}
Only one deviation indicator can be set per point of application:
\begin{flalign}
    & \sum_{j \in J} u_{c, p, j}^{\text{SquareSum}} = 1 & \forall c \in C, p \in c
\end{flalign}
Link variables $s_{c, p, d}$ and $u_{c}^{\text{StepSum}}$:
\begin{flalign}
    & M \cdot u_c^{\text{StepSum}} \geq s_{c, p, d} & \forall c \in C, p \in c, d \in p
\end{flalign}
A subevent is assigned exactly one starting time and the number of assigned resources equals the number of event resources ($|er|_{se}$):
\begin{flalign}
    & \sum_{t \in T, r \in er} x_{se, t, er, r} = 1 & \forall se \in SE, er \in se \\
    & \sum_{er \in se, r \in er} x_{se, t, er, r} = |er|_{se} \cdot y_{se,t} & \forall se \in SE, t \in T 
\end{flalign}
Link variables $v_{t, r}$ and $w_{se, er, r}$:
\begin{flalign}
    & \sum_{se \in SE, er \in se, t' \in T_{se, t}^{\text{start}}} x_{se, t', er, r} = v_{t, r} & \forall t \in T \backslash \{t_D\}, r \in R \\
    & \sum_{t \in T} x_{se, t, er, r} = w_{se, er, r} & \forall se \in SE, er \in se, r \in er 
\end{flalign}
The following new constraints link variables $w_{se, er, r}$ and our new variables $b_{e, r}$:
\begin{flalign}
    & w_{se, er, r} \leq b_{e, r} & \forall e \in E, se \in e, er \in e, r \in R \\
    & \sum_{se \in e} w_{se, er, r} \geq b_{e, r} & \forall e \in E, er \in e, r \in R
\end{flalign}
To link our new variables $c_{eg, r}$ to $b_{e, r}$ we need two more new constraints:
\begin{flalign}
    & b_{e, r} \leq c_{eg, r} & \forall eg \in EG, e \in eg \\
    & \sum_{e \in eg} b_{e, r} \geq c_{eg, r} & \forall eg \in EG
\end{flalign}
A subevent can not be assigned a start time that does not have enough times after it to fit its duration:
\begin{flalign}
    & y_{se, t} = 0 & \forall se \in SE, t \in T \backslash \{t_D\}, \rho(t) + D_{se} - 1 > |T|
\end{flalign}
Only a subset of the subevents are active at a time (since we create all possible subevents). A subevent is considered active if it has a starting time or resource assigned:
\begin{flalign}
    & \sum_{r \in er \backslash \{r_D\}} w_{se, er, r} \leq u_{se} & \forall se \in SE, er \in se, PA_{er} = 0 \\
    & \sum_{t \in T \backslash \{t_D\}} y_{se, t} \leq u_{se} & \forall se \in SE \\
    & \sum_{t \in T \backslash \{t_D\}} y_{se, t} + \sum_{\substack{r \in er \backslash \{r_D\}, \\ er \in se, PA_{er} = 0}} w_{se, er, r} \geq u_{se} & \forall se \in SE
\end{flalign}
The sum of the durations of the subevents of an event must equal the total duration of the event:
\begin{flalign}
    & \sum_{se \in e} D_{se} * u_{se} = D_e & \forall e \in E
\end{flalign}
Linking the variables $q_{r, t}$ and $p_{r, tg}$ that indicate if a resource is busy:
\begin{flalign}
    & |SE| \cdot q_{r, t} \geq v_{t, r} & \forall r \in R, t \in T \backslash \{t_D\} \\
    & q_{r, t} \leq v_{t, r} & \forall r \in R, t \in T \backslash \{t_D\} \\
    & p_{r, tg} \geq q_{r, t} & \forall r \in R, tg \in TG, t \in tg \\
    & p_{r, tg} \leq \sum_{t \in tg} q_{r, t} & \forall r \in R, tg \in TG\\
\end{flalign}
Events that have a given start time must have that time assigned ($se*$, represents an arbitrarily chosen subevent that has the same duration as the event $e$):
\begin{flalign}
    & y_{se*, e_{\text{Time}}} = 1 & \forall e \in E, e_{\text{Time}} \neq None, se*_{\text{Duration}} = e_{\text{Duration}}
\end{flalign}

\paragraph{Assign Resource Constraint}
\hfill \break
\hfill \break
\textbf{Applies to:} Events \\
\textbf{Point-of-application:} Event resource \\
\begin{flalign}
    & D_e - \sum_{\substack{se \in e, \\ r \in er \backslash \{r_D\}}} D_{se} \cdot w_{se, er, r} = s_{c, er}^{\text{assignres}} & \forall c \in \overline{C}, e \in c, er \in e, role_{er} = role_{c} 
\end{flalign}

\paragraph{Assign Time Constraint}
\hfill \break
\hfill \break
\textbf{Applies to:} Events \\
\textbf{Point-of-application:} Event\\
\begin{flalign}
    & D_e - \sum_{t \in T \backslash \{t_D\}, se \in e} D_{se} \cdot y_{se, t} = s_{c, e}^{\text{assigntime}} & \forall c \in \overline{C}, e \in c 
\end{flalign}

\paragraph{Split Events Constraint}
\hfill \break
\hfill \break
\textbf{Applies to:} Events \\
\textbf{Point-of-application:} Event\\
We use the parameters $\underline{B}_c^{\text{amount}}$ and $\overline{B}_c^{\text{amount}}$ to denote the minimum and maximum amount of events respectively. 
Likewise, we use the parameters $\underline{B}_c^{dur}$ and $\overline{B}_c^{dur}$ to denote the minimum and maximum duration of a subevent that is part of a given event, respectively.
The full deviation of a constraint $c \in \overline{C}$ is given by $s_{c, e}^{\text{spliteventamount}} + s_{c, e}^{\text{spliteventdur}}$
\begin{flalign}
    & U_{\underline{B}_c^{\text{amount}}, \overline{B}_c^{\text{amount}}} \sum_{se \in e} u_{se} \leq s_{c, e}^{\text{spliteventamount}} & \forall c \in \overline{C}, e \in c \\
    & \sum_{se \in e, \underline{B}_c^{dur} > D_{se}, \overline{B}_c^{dur} < D_{se}} u_{se} = s_{c, e}^{\text{spliteventdur}} & \forall c \in \overline{C}, e \in c
\end{flalign}

\paragraph{Distribute Split Events Constraint}
\hfill \break
\hfill \break
\textbf{Applies to:} Events \\
\textbf{Point-of-application:} Event\\
We use the parameters $\underline{B}_c$ and $\overline{B}_c$ to denote the minimum and maximum number of subevents respectively.
$D_c$ denotes the duration for which the constraint applies.
\begin{flalign}
    & U_{\underline{B}_c, \overline{B}_c} \sum_{se \in e, D_{se} = D_c} u_{se} \leq s_{c, e, er}^{\text{distsplitevent}} & \forall c \in \overline{C}, e \in c
\end{flalign}

\paragraph{Prefer Resources Constraint}
\hfill \break
\hfill \break
\textbf{Applies to:} Events \\
\textbf{Point-of-application:} Event resource \\
\begin{flalign}
    & \sum_{\substack{se \in e, r \notin c, \\ r \in R \backslash \{r_D\}}} D_{se} \cdot w_{se, er, r} = s_{c, er}^{\text{preferres}} & \forall c \in \overline{C}, e \in c, er \in e, PA_{er} = 0, role_{er} = role_{c} 
\end{flalign}

\paragraph{Prefer Times Constraint}
\hfill \break
\hfill \break
\textbf{Applies to:} Events \\
\textbf{Point-of-application:} Event\\
If $D_c$ is given only sub-events of Duration $D_c$ are considered. Otherwise, all sub-events are considered ($D_c = D_se$ is removed from sum).
\begin{flalign}
    & \sum_{\substack{se \in e, t \notin c, \\ t \in T \backslash \{t_D\}, D_c = D_se}} D_{se} \cdot y_{se, t} = s_{c, e}^{\text{prefertime}} & \forall c \in \overline{C}, e \in c 
\end{flalign}

\paragraph{Avoid Split Assignments Constraint}
\hfill \break
\hfill \break
\textbf{Applies to:} Event Groups \\
\textbf{Point-of-application:} Event Group\\
We slightly simplify this constraint compared to the original formulation, since we can make use of our new $c$ variables.
\begin{flalign}
    & \sum_{\substack{er \in e, PA_{er} = 0, \\ role_c = role_{er}}} w_{se, er, r} \leq k_{c, eg, r} & \forall c \in \overline{C}, r \in R, eg \in c, e \in eg, se \in e \\
    & \sum_{r \in R} k_{c, eg, r} - 1 \leq s_{c, eg}^{\text{avoidsplit}} & \forall c \in \overline{C}, eg \in c 
\end{flalign}

\paragraph{Spread Events Constraint}
\hfill \break
\hfill \break
\textbf{Applies to:} Event Groups \\
\textbf{Point-of-application:} Event Group\\
We use parameters $\underline{B}_{c, tg}$ and $\overline{B}_{c, tg}$ to denote the minimum and maximum number of sub-events of a given event that can be placed in time group $tg$ of a constraint $c \in \overline{C}$
\begin{flalign}
    & U_{\underline{B}_{c, tg}, \overline{B}_{c, tg}} \sum_{se \in e \in eg, t \in tg} y_{se, t} \leq s_{c, eg, tg}^{\text{spreadevent}} & \forall c \in \overline{C}, eg \in c, tg \in c
\end{flalign}

\paragraph{Link Events Constraint}
\hfill \break
\hfill \break
\textbf{Applies to:} Event Groups \\
\textbf{Point-of-application:} Event Group\\
We define the binary variable $o_{e, t}$ that takes the value 1 if at least one sub-event of event $e$ is scheduled at time $t$, and 0 otherwise.
We define the binary variable $l_{eg, t}$ that takes the value 1 if at least one event in event group $eg$ is scheduled at time $t$, and 0 otherwise.
\begin{flalign}
    & \sum_{t' \in T_{se, t}^{\text{start}}} y_{se, t'} \leq o_{e, t} & \forall e \in E, se \in e, t \in T \backslash \{t_D\} \\
    & \sum_{se \in e, t' \in T_{se, t}^{\text{start}}} y_{se, t'} \geq o_{e, t} & \forall e \in E, t \in T \backslash \{t_D\} \\
    & l_{eg, t} \geq o_{e, t} & \forall eg \in EG, e \in eg, t \in T \backslash \{t_D\} \\
    & l_{eg, t} - o_{e, t} \leq s_{c, eg, t}^{\text{linkevent}} & \forall c \in \overline{C}, eg \in c, e \in eg, t \in T \backslash \{t_D\}
\end{flalign}

\paragraph{Order Events Constraint}
\hfill \break
\hfill \break
\textbf{Applies to:} Pairs of Events \\
\textbf{Point-of-application:} Pair of Events\\
The variables $h_e^{\text{first}}$ and $h_e^{\text{last}}$ represent the first and last time assigned to any subevent of event $e$.
We use parameters $\underline{B}_{c}$ and $\overline{B}_{c}$ to denote the minimum and maximum number of times to separate the pair of events $(e, e')$ which are specified in constraint $c \in \overline{C}$.
\begin{flalign}
    & \rho (t) \cdot y_{se, t} + D_{se} \leq h_{e}^{\text{last}} & \forall c \in \overline{C}, e \in c, se \in e, t \in T \\
    & |T| - (|T| - \rho (t)) \cdot y_{se, t} \leq h_{e}^{\text{first}} & \forall c \in \overline{C}, e \in c, se \in e, t \in T \\
    & U_{\underline{B}_{c}, \overline{B}_{c}} (h_e^{\text{last}} - h_{e'}^{\text{first}}) \leq s_{c, (e, e')}^{\text{orderevent}} & \forall c \in \overline{C}, (e, e') \in c
\end{flalign}

\paragraph{Avoid Clashes Constraint}
\hfill \break
\hfill \break
\textbf{Applies to:} Resources \\
\textbf{Point-of-application:} Resource \\
\begin{flalign}
    & v_{t, r} - 1 \leq s_{c, r, t}^{\text{avoidclashes}} & \forall c \in \overline{C}, r \in c, t \in T \backslash \{t_D\}
\end{flalign}

\paragraph{Avoid Unavailable Times Constraint}
\hfill \break
\hfill \break
\textbf{Applies to:} Resources \\
\textbf{Point-of-application:} Resource \\
\begin{flalign}
    & \sum_{t \in c} q_{r, t} = s_{c, r}^{\text{unavailabletimes}} & \forall c \in \overline{C}, r \in c
\end{flalign}

\paragraph{Limit Idle Times Constraint}
\hfill \break
\hfill \break
\textbf{Applies to:} Resources \\
\textbf{Point-of-application:} Resource \\
We define the binary variables $h_{r, tg, t}^{\text{before}}$ and $h_{r, tg, t}^{\text{after}}$ to indicate if any events are happening before and after time $t$ in the timegroup $tg$, respectively.
We define the binary variables $h_{r, tg, t}^{\text{timeslot}}$ to indicate if time t is an idle time.
We define the integer variable $h_{r, tg}^{\text{timegroup}}$ to indicate the total amount of idle times in timegroup $tg$
We use $|tg|$ to indicate the amount of times in a time group $tg$.
We also use parameters $\underline{B}_{c}$ and $\overline{B}_{c}$ to denote the minimum and maximum values specified in constraint $c \in \overline{C}$.
\begin{flalign}
    & q_{r, t_2} \leq h_{r, tg, t_1}^{\text{before}} & \forall r \in C_R, tg \in C_{TG}, t_1, t_2 \in tg, \rho(t_1) > \rho(t_2) \\
    & \sum_{t_2 \in tg, \rho(t_1) > \rho(t_2)} q_{r, t_2} \geq h_{r, tg, t_1}^{\text{before}} & \forall r \in C_R, tg \in C_{TG}, t_1 \in tg \\
    & q_{r, t_2} \leq h_{r, tg, t_1}^{\text{after}} & \forall r \in C_R, tg \in C_{TG}, t_1, t_2 \in tg, \rho(t_1) < \rho(t_2) \\
    & \sum_{t_2 \in tg, \rho(t_1) < \rho(t_2)} q_{r, t_2} \geq h_{r, tg, t_1}^{\text{after}} & \forall r \in C_R, tg \in C_{TG}, t_1 \in tg \\
    & h_{r, tg, t}^{\text{before}} - q_{r, t} + h_{r, tg, t}^{\text{after}} - 1 \leq h_{r, tg, t}^{\text{timeslot}} & \forall r \in C_R, tg \in C_{TG}, t \in tg \\
    & -q_{r, t} + 1 \geq h_{r, tg, t}^{\text{timeslot}} & \forall r \in C_R, tg \in C_{TG}, t \in tg \\
    & h_{r, tg, t}^{\text{before}} \geq h_{r, tg, t}^{\text{timeslot}} & \forall r \in C_R, tg \in C_{TG}, t \in tg \\
    & h_{r, tg, t}^{\text{after}} \geq h_{r, tg, t}^{\text{timeslot}} & \forall r \in C_R, tg \in C_{TG}, t \in tg \\
    & \sum_{t \in tg} h_{r, tg, t}^{\text{timeslot}} = h_{r, tg}^{\text{timegroup}} & \forall r \in C_R, tg \in C_{TG} \\
    & U_{\underline{B}_c, \overline{B}_c} \sum_{tg \in c} h_{r, tg}^{\text{timegroup}} \leq s_{c,r}^{\text{idletimes}} & \forall c \in \overline{C}, r \in c 
\end{flalign}

\paragraph{Cluster Busy Times Constraint}
\hfill \break
\hfill \break
\textbf{Applies to:} Resources \\
\textbf{Point-of-application:} Resource \\
We use parameters $\underline{B}_{c}$ and $\overline{B}_{c}$ to denote the minimum and maximum values specified in constraint $c \in \overline{C}$.
\begin{flalign}
    & U_{\underline{B}_c, \overline{B}_c} \sum_{tg \in c} p_{r, tg} \leq s_{c, r}^{\text{clusterbusy}} & \forall c \in \overline{C}, r \in c 
\end{flalign}

\paragraph{Limit Busy Times Constraint}
\hfill \break
\hfill \break
\textbf{Applies to:} Resources \\
\textbf{Point-of-application:} Resource \\
We use parameters $\underline{B}_{c}$ and $\overline{B}_{c}$ to denote the minimum and maximum values specified in constraint $c \in \overline{C}$.
\begin{flalign}
    & -|tg| \cdot (1 - p_{r, tg}) + U_{\underline{B}_c, \overline{B}_c} \sum_{t \in tg} q_{r, t} \leq s_{c, r, tg}^{\text{limitbusy}} & \forall c \in \overline{C}, r \in c, tg \in c
\end{flalign}

\paragraph{Limit Workload Constraint}
\hfill \break
\hfill \break
\textbf{Applies to:} Resources \\
\textbf{Point-of-application:} Resource \\
The workload of a solution resource is given by $w_{e, se, er} = \frac{D_{se} \cdot L_{er}}{D_e}$ where $L_{er}$ is an integer denoting the workload of event resource er.
We use parameters $\underline{B}_{c}$ and $\overline{B}_{c}$ to denote the minimum and maximum values specified in constraint $c \in \overline{C}$.
\begin{flalign}
    & U_{\underline{B}_c, \overline{B}_c} \sum_{e \in c, t \in T \backslash \{t_D\}, se \in e, er \in e} w_{e, se, er} \cdot x_{se, t, er, r} \leq s_{c, r}^{\text{limitworkload}} & \forall c \in \overline{C}, r \in c 
\end{flalign}

\paragraph{Balance Class Size Constraint}
\hfill \break
\hfill \break
\textbf{Applies to:} Event Groups \\
\textbf{Point-of-application:} Event Group \\
We use the parameter $\overline{B}_{c}$ to denote the maximum class size difference specified in constraint $c \in \overline{C}$.
The role parameter is optional and denotes that only resources of a specific role in the event should be considered
We use the variable $mr_{c, eg}$ to denote the number of resources allocated to the specified Event Group
\begin{flalign}
    & \sum_{\substack{e \in eg, er \in e, \\ r \in er, type_{r} = type_{c} \backslash \{r_D\}}} c_{eg, r} = mr_{c, eg} & \forall c \in \overline{C}, eg \in \overline{C} \\
    & mr_{c, eg} - mr_{c, eg2} - \overline{B}_{c} \leq s_{c, eg}^{\text{balancesize}} & \forall c \in \overline{C}, eg \in c, eg2 \in c, eg \neq eg2 \\
    & mr_{c, eg2} - mr_{c, eg} - \overline{B}_{c} \leq s_{c, eg}^{\text{balancesize}} & \forall c \in \overline{C}, eg \in c, eg2 \in c, eg \neq eg2
\end{flalign}

\paragraph{Student Choice Constraint} 
\hfill \break
\hfill \break
\textbf{Applies to:} Resources \\
\textbf{Point-of-application:} Resource\\
We use parameters $\underline{B}_{c}$ and $\overline{B}_{c}$ to denote the minimum and maximum values specified in constraint $c \in \overline{C}$.
\begin{flalign}
    & U_{\underline{B}_c, \overline{B}_c} \sum_{eg \in c, e \in eg, er \in e, r \in er, r = r_c} c_{eg, r} \leq s_{c, r}^{\text{studentchoice}} & \forall c \in \overline{C}, r_c \in c
\end{flalign}

\subsubsection{Further Additions for guaranteeing exact objective values}
\hfill \break
\hfill \break
Using the model as described above will result in valid/optimal schedules if the ILP solver is run without a time limit. If there is a time limit the objective value
provided by the ILP solver might be bigger than the actual objective value of the produced schedule. This is due to the formulation using the $\leq$ operator in combination
with the deviation variables. So while it does lead to a worse objective value the ILP solver can save computational time by setting the deviation variables higher than necessary.
To mitigate this issue for the constraints that feature an upper and lower bound we can add binary indicators that tell the ILP solver which constraint to use based on the 
current assignments (e.g. if the constraints have the form $U_{\underline{B}_c, \overline{B}_c} x \leq s$) we set a binary variable $bin_{max} = 1$ if $x > \overline{B}_c$ and a
binary variable $bin_{min} = 1$ if $x < \underline{B}_c$. We then use those indicators to add the following constraints:

\begin{flalign}
    & x - \overline{B}_c = s & \text{ if } bin_{max} \\
    & \underline{B}_c - x = s & \text{ if } bin_{min}
\end{flalign}

We proceed similarly for constraints of the form $x_1 \leq s$, $x_2 \leq s$, \dots, $x_n \leq s$ (where $x_i$ may represent an arbitrary linear expression). We can instead model them with the following constraint:

\begin{flalign}
    & \max (x_1, \dots, x_n) = s &
\end{flalign}

Note that while we use so-called general constraints (indicator constraints, max constraints etc.) the ILP solver automatically transforms those into a set of linear constraints.

\end{document}